# Counterfactual Reasoning, Realism and Quantum Mechanics: Much Ado About Nothing?

Federico Laudisa

Department of Human Sciences, University of Milan-Bicocca
Piazza dell'Ateneo Nuovo 1, 20126 Milan, Italy

**Abstract**

I purport to show why old and new claims on the role of counterfactual reasoning for the EPR argument and the Bell theorem are unjustified: once the logical relation between locality and counterfactual reasoning is clarified, the use of the latter does no harm and the nonlocality result can well follow from the EPR premises. To show why, I critically review (i) incompleteness arguments that Einstein developed before the EPR paper, and (ii) more recent claims that equate the use of counterfactual reasoning with the assumption of a strong form of realism.

## 1 Introduction

The debates on the EPR argument and the status of nonlocality, which still prove to be among the most hotly discussed in the foundations of physics today, continue to reserve a special place for that special form of reasoning that has been called 'counterfactual'. At first sight, this appears hardly surprising given the role of counterfactual reasoning for scientific thought in general. If for instance we focus on the role that natural laws play in scientific theories, we realize that – whatever view of laws we might have – an essential task of laws themselves is to set boundaries for what is 'possible' and 'impossible' in nomological terms, namely for what might or might not happen compatibly with the laws and, consequently, for what *might have happened* (even if, in fact, it did not). More in detail, however, an option frequently adopted since the original formulation of the EPR argument has been to claim that counterfactual reasoning was an essential ingredient of the Bell theorem and to reject as a consequence the generality of the Bell theorem, due to the role that such form of reasoning is supposed to play when the very premises of the EPR argument itself are at work. Since the use of counterfactual reasoning was always taken in itself as an assumption at the outset of a form of realism that quantum mechanics cannot *in principle* support, then – so the argument goes – we can legitimately conclude that the Bell theorem need not imply that any theory preserving the predictions of quantum mechanics must be nonlocal. The idea that the Bell theorem, far from showing the microworld to be nonlocal, would be nothing but a further proof that "unperformed experiments have no results" (Peres 1978) is widespread, to such an extent that even many of those who take seriously nonlocality as a property of natural processes at the microlevel take for granted that in order to prove the Bell theorem we need assuming that unperformed experiments *do* have results. In a fairly recent paper, for instance, Stefan Wolf proposed an algorithmic-complexity view of nonlocality exactly with the motivation that it need not rely on the so-called counterfactual definiteness, taken by the author to be a standard assumption of the Bell theorem in any of its variants. According to Wolf "nonlocal correlations are usually understood through the outomes of alternative measurements (on two or more parts of a system) that cannot altogether actually be carried about in an experiment. Indeed, a joint input-output — e.g., measurement-setting–outcome — behavior is nonlocal if and only if the outputs for *all* possible inputs cannot coexist consistently. It has been argued that this counterfactual view is how Bell's inequalities and their violations are to be seen. I propose an alternative perspective which refrains from setting into relation the results of mutually exclusive measurements, but that is based solely on data actually available" (Wolf 2015, 052102-1, emphasis in the original) and in the sequel the author makes the point clear: "Nonlocal correlations are a fascinating feature of quantum theory. […] More specifically, the difficulty manifests itself when alternatives are taken into account: the argument leading up to a Bell inequality relates outcomes of different measurements, only *one* of which can actually be realized. […] Nonlocality is the impossibility of the outputs to *alternative inputs* to consistently coexist. Such reasoning assumes and concludes that certain pieces of *classical information exist or do not exist*." (Wolf 2015, 052102-1, emphasis in the original).[1]

In what follows, I purport to show why these claims on the role of counterfactual reasoning for the EPR argument and the Bell theorem are unjustified: once the logical relation between locality and counterfactual reasoning is clarified, the use of the latter does no harm and the nonlocality result can well follow from the EPR premises. In the section 2, which is more historically oriented, I critically review the role of counterfactual reasoning in EPR-like contexts, although with a special emphasis on incompleteness arguments that Einstein developed *before* the EPR paper. In the section 3 I briefly address the issue of the Einstein-Bell relation on the basis of the discussion conducted in the previous section, whereas in the section 4 I turn to a fairly recent attempt to equate the use of counterfactual reasoning with the assumption of a strong form of realism: here I explore the reasons why this attempt – along with the many that preceded it – fails to be convincing. The final section draws some general conclusions.

---

[1] A side remark. Given a thesis X, it is quite peculiar to say that a certain argument *both* 'assumes' and 'concludes' X: the logical status of X within the argument – hence, the argument itself – changes dramatically if X is an assumption or the outcome of a derivation! As we will see later, this is exactly the heart of the matter with counterfactual definiteness in EPR-Bell kind of arguments.

## 2  On the role of counterfactual reasoning in pre- and post-EPR incompleteness arguments for quantum mechanics?

Although the EPR paper occupied the center of stage for many years, it is now well known that the formulation of the argument therein is only one of several formulations that Einstein provided for his quantum incompleteness claim between 1927 and the last years of his life. In general terms, from the thought experiment proposed at the Solvay Conference in 1927 onward, his focus on the peculiar features of the state representation in quantum mechanics highly contributed *in retrospect* to realize how crucial is entanglement to the theory in so many respects[2]. But also for our purposes the alternative formulations of an incompleteness argument turn out to be especially useful, since they show that it is locality rather than any counterfactual, pre-existence claim that is bound to play the fundamental role in highlighting a disconcerting feature of quantum mechanics.

As is again well known, the main sources of misunderstanding surrounding the EPR argument in its original form are two: the first is the infamous reality condition and the second is the choice of two conjugate observables like position and momentum as the relevant observables. The latter proved especially misleading as far as the Bohr response to EPR was concerned: Bohr in fact questioned the argument on the basis of a somewhat obscure kind of disturbance theory of measurement[3], while the ingenuity of the thought experiment was exactly that of avoiding altogether any assumption that might imply any disturbance of a quantum sort[4].

---

[2] This is one of the episodes in the history of contemporary physics where the received view of Einstein as a reactionary scientist, unable to follow the more promising lines of development of the new physics in the second half of his life, proves so dramatically inadequate. As an instance of this inadequacy, let me recall what Werner Heisenberg said in his introduction to *The Born-Einstein Letters*, commenting on the Einstein's attitude toward quantum mechanics: "Most scientists are willing to accept new empirical data and to recognize new results, provided they fit into their philosophical framework. But in the course of scientific progress it can happen that a new range of empirical data can be completely understood only when the enormous effort is made to enlarge this framework and to change the very structure of the thought processes. In the case of quantum mechanics, Einstein was apparently no longer willing to take this step, or perhaps no longer able to do so" (Born 2005, xxxvii).
[3] This theory is downgraded by Arthur Fine to the status of a 'semantic' theory of disturbance (Fine 1986, 35).
[4] See the deep analysis of this point in the chapter 3 of Fine 1986; see also Maudlin 2014, 424010-12,13.

In any case, it was eight years before EPR that Einstein first proposed a thought experiment in which the dilemma at the heart of the EPR argument – namely completeness vs. locality – is spelled out with no reference to unclear 'elements of reality', 'counterfactual definiteness' and so on.

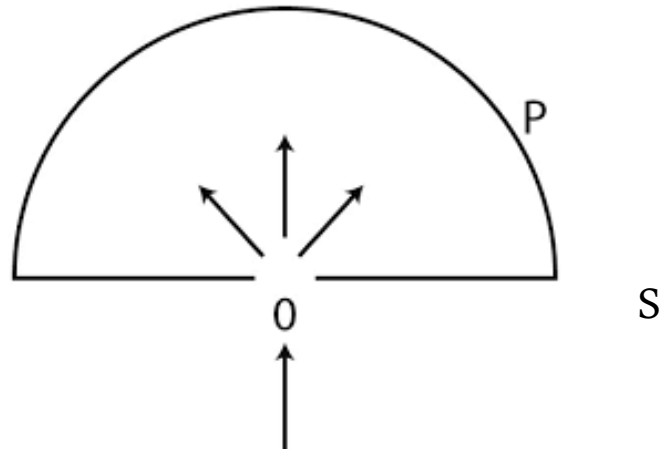

– **Figure 1** –

The thought experiment contemplates a diaphragm S, with an opening O, and a hemispherical detection screen P (Figure 1): incident particles indicated by the arrow typically are expected to diffract and produce a wave propagating toward P. If a single electron is sent toward the opening, at the end of the path it will be detected at some specific point on the screen with the probability prescribed by the wave function $\psi$ associated with the pre-detection state of the electron. According to Einstein, we might account for the situation in two possible, mutually exclusive ways:

> *Conception I* – The de Broglie-Schrödinger waves do not correspond to a single electron, but to a cloud of electrons extended in space. The theory gives no information about individual processes, but only about the ensembles of an infinity of elementary processes.
>
> *Conception II* – The theory claims to be a complete theory of individual processes.
>
> […] According to the first, purely statistical point of view $|\psi|^2$ expresses the probability that there exists at the point considered a *particular* particle of the cloud, for example at a given point on the screen. According to the second, $|\psi|^2$ expresses the probability that at a given instant *the same* particle is present at a given point (for example on the screen). Here, the theory refers to an individual process and claims to describe everything that is governed by laws.

After the presentation of the two options, between which Einstein clearly favours the first, he focuses on an especially awkward consequence of the second:

> If $|\psi|^2$ were simply regarded as the probability that at a certain point a given particle is found at a given time, it could happen that *the same* elementary process produces an action *in two or several* places on the screen. But the interpretation according to which $|\psi|^2$ expresses the probability that *this* particle is found at a given point, assumes an entirely peculiar mechanism of action at a distance, which prevents the wave continuously distributed in space from producing an action in *two* places on the screen. [...] If one works solely with the Schrödinger waves, interpretation II of $|\psi|^2$ implies to my mind a contradiction with the postulate of relativity. (Bacciagaluppi, Valentini 2009, 441).

The negation of "an entirely peculiar mechanism of action at a distance", which in the Einstein view follows straight from relativity, is exactly what we call "locality": *if* we assume it, *then* we decide to accept the good old intuition according to which there is a matter of fact concerning the pre-detection properties of particle. Since the ψ cannot in principle account for such properties, it follows that the quantum-mechanical description of the pre-detection state of the particle is incomplete. Full stop. No 'elements of reality', no 'counterfactual definiteness', no simultaneous values for incompatible observables, nothing.

A slightly different formulation, inspired by the argument Einstein proposed in 1927, is known as the *Einstein boxes* experiment and supposed to emphasize even more dramatically the dilemma between locality and completeness[5].

(i)

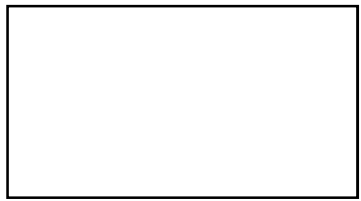

(ii)

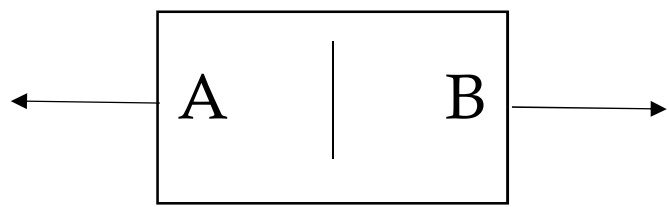

(iii)

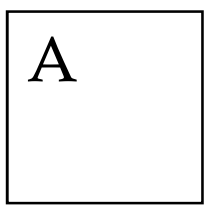 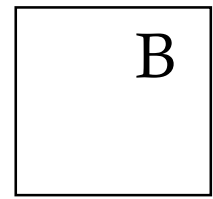

---
[5] See Hardy 1998, 422, Norsen 2005, 164-165.

– **Figure 2** –

Let us suppose (Figure 2) first we have a box with a particle in it (i). Let us suppose then that we realize a partition (ii) into two different boxes – that we call A and B – that we move to regions that are far separated from each other (iii). According to quantum mechanics, the state of the system at this stage can be written as

$$1/\sqrt{2}\,[\psi_A + \psi_B]$$

where $\psi_A$ ($\psi_B$) represents that the particle is in box A (B). Now let us open the box A and suppose we find the particle there, we know that certainly the particle will not be found in the box B. Two possible options for explaining this fact, however, are available:

(1) either we accept that the particle was definitely in box A (so that opening the box simply amounts to discover the actual, pre-detection state of affairs),

(2) or we accept that the act of opening the box coincides with the fact that the particle is in box A (so that there was no actual, pre-detection state of affairs concerning where the particle was).

If we choose the option (1), then quantum mechanics is incomplete since the state description $1/\sqrt{2}\,[\psi_A + \psi_B]$ cannot account for the actual, pre-detection state of affairs that the particle is in box A; if on the other hand we choose (2), and accept as a consequence that the quantum state description is complete, then the information that the particle cannot also be found in box B, since it has been found in box A, can be conveyed only by a non-local influence. I wish to stress that if we assume locality, (1) is bound to follow: in this case, the fact that the particle is in box A – in spite of the state of the system being described by $1/\sqrt{2}\,[\psi_A + \psi_B]$ – *is justified* by denying the possibility that the mere act of opening a box could cause nonlocally the particle to collapse onto the state 'being in box A'. In other terms, option (1) is *not in itself* the option of assuming the existence of an actual, pre-detection state of affairs *over and above* the quantum state description; it is rather the option according to which the existence of an actual, pre-detection state of affairs over and above the quantum state description is allowed *on the basis of locality*[6].

A further, simplified version of the Einstein 1927 thought experiment, due to Hardy 1995, is even more effective in showing why locality is the key issue and why counterfactual reasoning does not support any strongly realistic assumption in its own right.

---

[6] This kind of thought experiment is reformulated by Einstein in the letter to Schrödinger of June 19th 1935 in which he complains about the published version of the EPR argument: see the discussion of it in Fine 1986, 35-38.

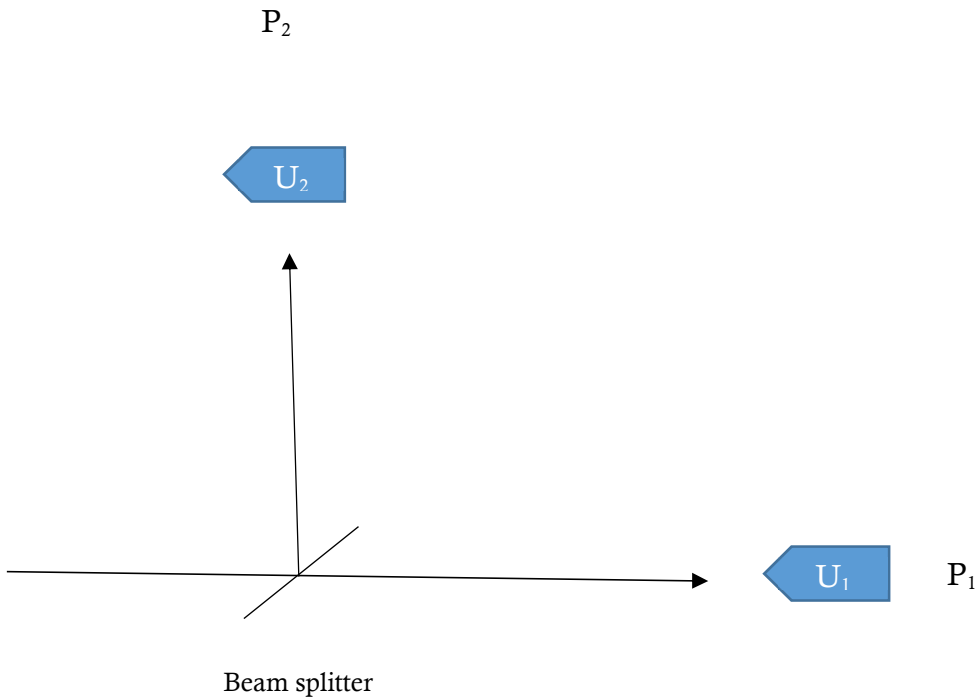

– **Figure 3** –

Let us consider a single particle incident on a beam splitter (Figure 3), where $P_1$ and $P_2$ denote the points where two detectors $U_1$ and $U_2$ respectively might be placed, distant from each other: clearly, each detector does or does not fire whether or not the particle is found at the respective point. If we put in fact a detector $U_1$ at $P_1$ (the role of $P_1,P_2$ and $U_1,U_2$ can be reversed of course), in principle we have to take into account two possibilities:

(a) either $U_1$ fires, which implies that a detector $U_2$, if put at $P_2$, *certainly would not fire* (denoted as $[U_2 = 0]$);

(b) or $U_1$ does not fire, which implies that a detector $U_2$, if put at $P_2$, *certainly would fire* (denoted as $[U_2 = 1]$).

This implies that, both in case (a) and (b), there is always a definite state of affairs concerning the particle at $P_2$. Moreover, if we assume locality, such state of affairs cannot depend on whether a detector $U_1$ was placed at $P_1$ or not. According to quantum mechanics, however, the pre-measurement state of the system is $1/\sqrt{2} \, [\psi_{[U1=1]} \, \psi_{[U2=0]} + \psi_{[U1=0]} \, \psi_{[U2=1]}]$: this state that cannot account for the definite state of affairs concerning the particle at $P_2$, therefore the quantum state description turns out to be incomplete. As should clear enough, it is locality that allows one to derive or justify the existence of a definite state of affairs concerning the particle at $P_2$. Moreover, it is still locality that makes the use of counterfactual reasoning innocuous and far from forcing any realism as an independent assumption: if we consider the counterfactual statement < *Should $U_1$ fire, $U_2$ certainly would not fire* >, it is the independence of the [not-firing]-state of affairs from whether a detector $U_1$ was placed at $P_1$ or not that makes it true, so that nowhere we are forced to assume the existence of [firing]- or [not-firing]-state of affairs *in themselves*.

## 3 Bell vs. Einstein?

A further, standard source of misunderstanding – which resonates with the misunderstandings we have discussed above – concerns the complex relationship between the Bell results and the presuppositions of the Einsteinian view of quantum mechanics. In what sense is this relationship complex and why can a more thorough investigation of it work effectively in both directions? For at least two, essential motivations. On the one hand, the Bell results are among the post-Einsteinian scientific achievements that most call into question some crucial features of the Einstein own image of the physical world. On the other hand, the Bell theorem itself turned out to be a sort of lens from which to look at some of the original Einstein's foundational views, a lens that more often than not produced serious distortions: not by chance, these distortions are often also distortions of the significance of the Bell results themselves.

If we focus in particular on the role that an implicit assumption of classically inspired form of realism is supposed to play since the first reservations of Einstein about quantum mechanics, the current representation of the Einstein-Bell connecting line is still not free of ambiguities. Some instances of this representation prove to be simply wrong:

> While a very tiny loophole in principle remains for local realism, it is a very safe position to assume that QM has definitely shown to be the right theory. Thus, a very deep question, namely whether or not events observed in the quantum world can be described by an underlying deterministic theory, has been answered by experiment, thanks to the momentous achievements of John Bell." (Bertlmann, Zeilinger 2002, viii).

The existence and mathematical consistency of a pilot-wave-theoretical formulation of quantum mechanics proves this statement to be just false: any de Broglie-Bohm type of theory, being both deterministic and non-local by construction, does not run afoul of the Bell theorem, which rules out only *local* completions – be they deterministic or indeterministic. Other statements, although closer to truth, are somewhat unclear. In his introduction to the second edition (2004) of the celebrated Bell collection *Speakable and Unspeakable in Quantum Mechanics*, Alain Aspect – a leading figure in the experimental work on the foundations of quantum mechanics and the Bell inequalities – for instance writes:

> John Bell demonstrated that there is no way to understand entanglement in the framework of the usual ideas of a physical reality localized in space-time and obeying to causality. This result was the opposite to the expectations of Einstein, who had first pointed out, with his collaborators Podolsky and Rosen, the strong correlations between entangled particles, and analyzed these correlations in the framework of ideas of a local physical reality. The most remarkable feature of Bell's work was undoubtedly the possibility it offered to determine *experimentally* whether or not Einstein's ideas could be kept (Bell 2004, p. xvii, emphasis in the original).

The conjunction of 'physical reality localized in space-time' and 'causality' on one side and the expression 'local physical reality' on the other lead us to think that on the background the following presupposition is lurking: namely, that the Bell theorem rules out completions of quantum mechanics that are both local *and* 'realistic' and that the latter is the Einstein's heritage that the theorem itself forces us to abandon. The point is crucial, since according to some the assumption of realism – being quantum-mechanically untenable – undermines the Bell theorem as a *non-locality* theorem.

As far as Einstein is concerned, I think it can be safely said that Einstein would have longed for a formulation of quantum mechanics able to include what might be called a *Pre-Existence* condition: a standard physical system has at all times a whole bunch of pre-existing properties – encoded in its state – no matter whether there is a measurement interaction or not. As we attempted to show in the previous section, however, the personal Einstein preference for such a hypothetical formulation does *not* imply *per se* that the Einstein incompleteness arguments – the pre-EPR ones, the EPR itself and the post-EPR ones – *do need* to require independently such pre-existence condition. Pre-Existence and locality interact in an interesting way in the following statement, taken from the Einstein *Autobiographical Notes*:

> There is to be a system which at the time *t* of our observation consists of two partial systems $S_1$, and $S_2$, which at this time are spatially separated [....] The total system is to be completely described through a known ψ-function $ψ_{12}$ in the sense of quantum mechanics. All quantum theoreticians now agree upon the following: If I make a complete measurement of $S_1$, I get from the results of the measurements and from $ψ_{12}$ an entirely definite ψ-function $ψ_2$ of the system $S_2$. The character of $ψ_2$ then depends upon what kind of measurement I undertake on $S_1$. Now it appears to me that one may speak of the real factual situation of the partial system $S_2$. Of this real factual situation, we know to begin with, before the measurement of $S_1$, even less than we know of a system described by the ψ-function. But on one supposition we should, in my opinion, absolutely hold fast: the real, factual situation of the system $S_2$ is independent of what is done with the system $S_1$ which is spatially separated from the former. (Einstein 1949, 85).

What Einstein calls 'the real, factual situation' is clearly supposed to be a pre-existent feature of the system, but the assumption that 'we should hold fast' – i.e. the independence of $S_2$ from what is done with the system $S_1$ and viceversa – is what *logically* guarantees that the system can be conceived as being in a well-defined real, factual situation at all times. Although the foundational views of Einstein show some ambivalence on whether pre-existence and locality are both fundamental assumptions or not, all his incompleteness arguments can be formulated with a clear logical relation between them, according to which the latter is a pre-requisite for the former[7].

It is ironic to note that many misunderstandings of the Bell work derive *exactly* from attributing to Bell himself the endorsement of some sort of the pre-existence claim. Here the misunderstanding could not more complete. It was already in his paper «On the problem of hidden variables in QM» (published in 1966 but written in 1963, *before* the paper that derives for the first time the Bell inequality) that Bell remarked how unreasonable it was to require pre-existence. With reference to the 'no-hidden-variable' proofs provided in the preceding years by von Neumann, Gleason and Jauch-Piron, Bell wrote:

> [...] these demonstrations require from the hypothetical dispersion free states not only that appropriate ensembles thereof should have all measurable properties of quantum mechanical states, but certain other properties as well. These additional demands appear reasonable when results of measurements are loosely identified with properties of isolated systems. They are seen to be quite unreasonable when one remembers with Bohr 'the impossibility of any sharp distinction between the behaviour of atomic objects and the interaction with measuring instruments which serve to define the conditions under which the phenomena appear'. (Bell 2004, 2).

It is (the first instance of) what Abner Shimony called a «judo-like manoeuvre» (Shimony 1984, 121): to use the Bohr *dictum* itself in order to show how implausible the proofs by von Neumann and the others turned out to be, *exactly because they did assume something like the pre-existence claim*.

## 4 Counterfactual realism: it ain't necessarily so…

---

[7] On the nature of Einstein's realism, see e.g. Howard 1993, Home-Whitaker 2007, ch. 8.

In spite of the arguments above, the controversy on an alleged realism assumption at the source of the Einstein-Bell arguments is still alive. I will now concentrate the discussion on a proposal about the interpretation of the Bell theorem, due to Marek Zukovski and Caslav Brukner (ZB, from now on). Focusing on their formulation is useful, since they touch two recurrent points in the debate on the role of counterfactual reasoning and they do it systematically, in a way that is especially suitable to a clean discussion. ZB concentrate their critical analysis on what they take to be two distinct views, simply called view (A) and view (B):

• (A) The derivation of a Bell inequality relies only on the premise of locality, and nothing more.

• (B) The derivation of a Bell inequality does rely on the additional premise of determinism, or any of the notions listed in the abstract. However, these notions can be derived from the premises of locality, the freedom of an experimenter to choose the setting of his/her local apparatus and quantum predictions and nothing more. (Zukovski & Brukner 2014, 424009-2)

The alleged refutations by ZB of both views (A) and (B) share the claim that, in both cases, what Bell called *local causality* is in fact "a *compound* condition" (Zukovski & Brukner 2014, 424009-2, emphasis in the original), namely the conjunction of locality *plus* an additional condition:

> as to (A), the additional condition would be "the assumption of the existence of a (positive and normalized) joint probability distribution for the values of all possible measurements that could be performed on an individual system, no matter whether any measurement – and which measurement – is actually performed" (as we will see later, the reference to actual-or-possible measurement already hints at the role of counterfactual reasoning);

> as to (B), the additional condition would be *counterfactual definiteness*, i.e. a condition that "allows one to assume the definiteness of the results of measurements, which were actually not performed on a given individual system. They are treated as unknown, but in principle *defined* values." (Zukovski & Brukner 2014, 424009-2, emphasis in the original).

After all both the assumptions that for ZB are implicit in view (A) and (B) respectively and that, conjoined with locality, make up the crucial, compound condition of the Bell theorem seem to point at some sort of Classicality that the theorem would be about: therefore, the argument goes, the joint outcome of both refutations is the conclusion that the Bell theorem, far from demonstrating nonlocality for the quantum world, would be just a sophisticated way of stating that the quantum world is not 'classical' (or, more modestly, that quantum mechanics is not a 'classical' theory. Let me review, then, the arguments about (A) and (B) in turn in the next two subsections.

## 4.1 View (A): locality and joint probability distributions

In the standard EPR-Bell context we suppose two parties, Alice and Bob, that we may associate with two different categories of events: the choice of a measurement setting and the production of an outcome. If we denote by $x$, $y$ two possible settings and by $A$, $B$ two possible outcomes respectively, we are generally interested to correlations described by distributions $p(A, B \mid x, y)$. In addition, it is customary to formulate a statistical framework in which these expressions like $p(A, B \mid x, y)$ represent distributions of more general states $\lambda$, such that

$$p(A, B \mid x, y) = \int d\lambda \rho(\lambda) \, p(A, B \mid x, y, \lambda) \qquad [1]$$

with $\rho(\lambda)$ being a suitable distribution over the 'general' state space $\Lambda$. Now, from the formal theory of probability we know that

$$p(A, B \mid x, y) = p(A \mid B, x, y) \, p(B \mid x, y),$$

so that

$$p(A, B \mid x, y) = \int d\lambda \rho(\lambda) \, p(A \mid B, x, y, \lambda) \, p(B \mid x, y, \lambda); \qquad [2]$$

moreover, locality requires that the outcomes $A$ and $B$ depend only on respective near-by causal factors, therefore [2] becomes

$$p(A, B \mid x, y) = \int d\lambda \rho(\lambda) \, p(A \mid x, \lambda) \, p(B \mid y, \lambda). \qquad [3]$$

According to ZB, [3] allows to introduce a joint probability distribution along the following lines. If $A_x$ [$B_y$] denote the outcome of an $A$-[$B$-]measurement when the chosen setting is $x$ [$y$], and 1, 2 denote two possible choices for each setting, the desired joint distribution can be written as

$$p(A_{x=1}, A_{x=2}, B_{y=1}, B_{y=2} \mid \lambda) = p(A_{x=1} \mid \lambda) \, p(A_{x=2} \mid \lambda) \, p(B_{y=1} \mid \lambda) \, p(B_{y=2} \mid \lambda) \qquad [4]$$

If this is the formal account, the ZB foundational criticisms are two. The first point is that the very introduction of the general state structure $\Lambda$ amounts to the introduction of additional hidden parameters, $\lambda$, which are *not* present in quantum mechanics. The $\lambda$ […] can pop up under many guises such as, e.g., 'the physical state of the systems as described by any possible future theory', 'local beables', 'the real state of affairs', 'complete description of the state', etc. Since $\lambda$ do not appear in quantum mechanics, they are (good old) *hidden variables*. Anything on which one conditions probabilities which gives a different structure to formulas for probabilities than quantum mechanical formalism is a hidden variable *per se*. (Zukovski & Brukner 2014, 424009-3, emphasis in the original)

This wording is meant to suggest that the mere use of a Λ-statistical theory is nearly equivalent (or leads necessarily) to the independent assumption of pre-determined values for any relevant observable. The second point reinforces the first: the introduction of a joint probability "for all possible outcomes under all possible pairs of settings" presupposes that "[T]hese outcomes include, for a single run of a Bell experiment, the actual measured outcomes and on an equal footing those which *could have been potentially measured*." (Zukovski & Brukner 2014, 424009-4, my emphasis). ZB argue, then, that these two points jointly undermine view (A): in their argument both the above points allegedly show that Bell 'local causality' brings with itself a hidden assumption of 'pre-determination', a circumstance that makes 'local causality' a compound condition and that, as a consequence, is bound to change radically the interpretation of the Bell theorem as a nonlocality result.

As to the first point, I wish to stress that equating the mere use of a Λ-statistical theory with an assumption of definiteness of values for all observables – even those for which quantum mechanics prevents such definiteness – looks like a *petitio principii*. On the contrary, the introduction of a general framework like a Λ-statistical theory is relatively innocent and perfectly reasonable by a foundational point of view: provided that we already know that quantum mechanics is nonlocal (via entanglement), we wonder whether it might be seen as a 'fragment' of a more general theory which might recover the condition of locality at a 'higher' level (Bell 1975, reprinted in Bell 2004, 55). The λ states just need be general enough to encompass quantum states, but nowhere it is required that this sort of 'inclusive' feature is in itself equivalent to the assumption that the Λ-statistical theory must be a 'classical' theory in the sense of pre-determined values for any meaningful physical quantity.

As to the second point, I remark that under the *only* locality assumption we obtain the factorization in [4]. This is hardly surprising and is exactly what we would expect: if an outcome is sensitive to causal factors that may affect it only locally, the probability of each outcome is independent from the probability of any other outcome. Conclusion: the outcome probabilities factorize *as a consequence* of locality. By this point of view, in order for the Bell argument to go through we do not need to assume the existence of a total joint probability distribution for all possible values: it is just locality that justifies the factorization of the specific probabilities we are interested in. If, on the other hand, we follow ZB in assuming that the factorization in [4] is really the restriction of a total joint probability distribution, we are in fact *unnecessarily imposing* *from the outside* the existence of values for all observables. Should we push the above argument a bit further, we might then *ex absurdo* add any condition we like to locality – let me call it the Whatever-Condition – and claim that the Bell theorem does not show the quantum world to be nonlocal but rather that it does not satisfy the Whatever-Condition! On the other hand, it is equally unsurprising that, if I assume an equivalence between locality in the sense of [3] and the existence of a joint probability distribution for all values as if these values were already defined (no matter whether they are measured or not), I can go *from* the existence of a joint probability distribution *to* locality: if values are fixed, they are insensitive to whatever else takes place remotely and therefore locality obtains[8].

## 4.2  View (B): counterfactual reasoning, the EPR argument and all that

As recalled above, the ZB refutation of View (B) amounts to showing that

> any attempts to derive determinism, or hidden-variables, via an EPR argument are futile. The argument of EPR was based on an assumption, which is often not noticed or is forgotten: *counterfactual definiteness*. This assumption allows one to assume the definiteness of the results of measurements, which were actually not performed on a given individual system. They are treated as unknown, but in principle *defined* values. This is in striking disagreement with quantum mechanics and the complementarity principle. (Zukovski & Brukner 2014, 424009-2, emphasis in the original).

The lesson that we learn from the joint refutations of views (A) and (B), then, "is that Bell's theorem, in all its forms, tells us *not* what quantum mechanics *is*, but what quantum mechanics *is not*." (Zukovski, Brukner 2014, 424009-8, emphasis in the original). With respect to the refutation of view (B), the ZB stance is widely shared. In an extensive review on the Bell theorem appeared on the *American Journal of Physics* in 2010, Guy Blaylock claimed "that the minimal assumptions behind Bell's predictions are locality and 'counterfactual definiteness' (that we can postulate a single definite result from a measurement even when the measurement is not performed)" (Blaylock 2010, 111). In the same journal in 2013 Lorenzo Maccone – while providing "an extremely simple proof of Bell's inequality" – claims that "Bell's theorem can be phrased as 'quantum mechanics cannot be both local and counterfactual-definite' " (Maccone 2013, 854)[9], whereas Stephen Boughn most recently states: "Bell's theorem clearly makes use of counterfactual definiteness; his inequality involves the correlations of the spins of the two particles in each of two different directions that correspond to non-commuting spin components. *This use of counterfactuals is entirely appropriate because it is used to investigate a test for classical hidden variable theories*." (Boughn 2017, 648, my emphasis)[10].

---

[8] The standard reference for all arguments relying on the equivalence between factorizability and existence of a joint probability distribution is Fine 1982 (for fairly recent examples, see Zukovski, Brukner 2014, Wolf 2015): see also Mermin 1992, where the author writes: "The Bell theorem, although it does rely on the properties of a particular state, proves the non-existence of a joint distribution for a set of observables *required to have one by an apparently common-sense notion of physical locality*" (27, my emphasis). In addition to the argument defended above in the main text, it can be shown that the Fine deterministic hidden variable model implicitly embodies locality anyway (Shimony 1984, Laudisa 1996).

I recall here that the view (B) is the claim that determinism (or any cognate notion) can be derived from the premises of locality, the freedom of an experimenter to choose the setting of his/her local apparatus and quantum predictions and nothing more. In a compact form, ZB depict view (B) as

$$\text{Freedom \& QF \& Locality} \Rightarrow \text{Determinism}$$

Usually this implication is traced back to the very EPR argument but ZB claim with no further justification that this reconstruction

is wrong. The logical structure of the EPR argumentation about elements of reality is

$$\text{Freedom \& QF \& Locality \& Counterfactual Definiteness}$$
$$\Rightarrow \text{for specific observables elements of reality exist for the EPR state}$$

The thesis of the implication is true, provided one does not make unfounded generalization of it to *arbitrary* observables and *arbitrary* states. Indeed, for the EPR state and momentum and position observables ($P$ and $Q$), elements of reality seem to be a consistent notion. However this is arrived at by considering two situations of which only one can be the case in the given run of the experiment (measuring either $P$ or $Q$). This is counterfactual definiteness at work. (Zukovski & Brukner 2014, 424009-6, emphasis in the original).

The flaw in this argument is similar to the one discussed in the previous section, namely the supposition that the existence of the elements of physical reality is *added* on top with no other justification than to show that this 'classical-like' supposition cannot hold in a quantum world. If, on the other hand, we take seriously (i) the fact that it is the locality assumption that is prior, and (ii) that it is exactly this assumption that legitimates to talk of elements of physical reality corresponding to spin measurements along different axes, we realize that the above appeal to counterfactual reasoning does no harm. Take the EPR argument in the Bohm formulation. *If* the theory by which we try to account for EPR correlation is assumed to be local, *then* the choice of a certain axis for spin measurement *here* is totally unaffected by what is the chosen axis for spin measurement *there*, and this holds perfectly the same also in a counterfactual sense. Suppose in the actual world the experimenter has chosen the axis *z*: if her measuring operations cannot affect by definition what axis is chosen – and what outcome has been obtained – on the other side, this holds naturally for *whatever choice* and this makes it invalid to claim that in an ordinary EPR argument we have adopted a conterfactual definiteness assumption *in addition* to locality.

---

[9] A remark in sociology of science: since the *American Journal of Physics* is the intended journal of US physics teachers, it is an revealing sign that in two issues of the journal in two different years readers are pedagogically taught that a ghostly assumption like counterfactual definiteness does play an essential role in the EPR argument and in the ensuing Bell theorem!
[10] For older, notable statements in the same direction see e.g. Unruh 1999, 2001.

Let us consider the point in a specific formulation of the EPR argument in the Bohm version, in which we have a composite quantum system $S_1+S_2$ of a pair of spin-1/2 particles $S_1$ and $S_2$. The composite system is prepared at a time $t_0$ in the singlet state $\Psi$

$$\Psi = 1/\sqrt{2}\,(|1,+>_n |2,->_n - |1,->_n |2,+>_n ),$$

where **n** denotes a generic spatial direction. We take into account the measurements concerning the spin components along given directions, whose possible outcomes are only two (conventionally denoted by '+1' and '-1'). We assume also that the spin measurements on $S_1$ and $S_2$ are performed when $S_1$ and $S_2$ occupy two mutually isolated spacetime regions $R_1$ and $R_2$. According to QM, we know that if the state of $S_1+S_2$ at time $t_0$ is $\Psi$, then the (reduced) states of the subsystems $S_1$ and $S_2$ at time $t_0$ are respectively

$$\rho(1,\Psi)=1/2(\mathbf{P}_{|1,+>n} + \mathbf{P}_{|1,->n}), \qquad \textbf{(RS)}$$
$$\rho(2,\Psi)=1/2(\mathbf{P}_{|2,+>n} + \mathbf{P}_{|2,->n}),$$

('RS' stands for 'reduced states') so that, for any **n,**

$$\text{Prob}_{\rho(1,\Psi)}\,(\text{spin}_n\ \text{of}\ S_1 = +1) = \text{Prob}_{\rho(1,\Psi)}\,(\text{spin}_n\ \text{di}\ S_1 = -1) = 1/2$$
$$\text{Prob}_{\rho(2,\Psi)}\,(\text{spin}_n\ \text{di}\ S_2 = +1) = \text{Prob}_{\rho(2,\Psi)}\,(\text{spin}_n\ \text{di}\ S_2 = -1) = 1/2$$

Moreover, if we perform at a time *t* a spin measurument on $S_1$ along **n** with outcome +1 (–1), a spin measurement on $S_2$ along **n** at a time $t' > t$ will give with certainty the outcome – 1 (+1), namely for any **n** ('**AC**' stands for '**A**nti**C**orrelation')

$$\text{Prob}_\Psi\,[(\text{spin}_n\ \text{di}\ S_1 = +1)\ \&\ (\text{spin}_n\ \text{di}\ S_2 = -1)] = \qquad \textbf{(AC)}$$
$$\text{Prob}_\Psi\,[(\text{spin}_n\ \text{di}\ S_1 = -1)\ \&\ (\text{spin}_n\ \text{di}\ S_2 = +1)] = 1.$$

Let us suppose now to perform at time $t_1 > t_0$ a spin measurement on $S_1$ with outcome +1. Therefore, according to (**AC**), a spin measurement on $S_2$ along **n** at a time $t_2 > t_1$ will give with certainty the outcome – 1. Let us suppose now to assume the following condition:

**REALITY** – If, without interacting with a physical system *S*, we can predict with certainty - or with probability 1 - the value **q** of a quantity **Q** pertaining to *S*, then **q** represents an objective property of *S* (denoted by [**q**]).

Then, for $t_2 > t_1$ [**spin** $_n$ = −1] represents an objective property of $S_2$. But might the objective property [**spin** $_n$ = −1] of $S_2$ have been somehow "created" by the spin measurement on the distant system $S_1$? The answer is negative if we assume the following condition:

**LOCALITY** – No objective property of a physical system *S* can be influenced by operations performed on physical systems that are isolated from *S*.

At this point, **LOCALITY** allows us to state the existence of the objective property [**spin** $_n$ = −1] for the system $S_2$ also at a time $t'$ such that $t_0 > t' > t_1$. Namely, if we assume that the measurement could not influence the validity of that property at that time, it follows that the property *was holding already at time t'*, a time that *precedes* the measurement performed on the other subsystem. But at time $t'$ the state of $S_1+S_2$ is the singlet state Ψ, therefore according to (**RS**) the state of $S_2$ is the reduced state $\rho(2,\Psi)=1/2(\mathbf{P}_{|2,+> n} + \mathbf{P}_{|2,-> n})$, that prescribes for the property [**spin** $_n$ = −1] of $S_2$ only a probability 1/2. Let us consider finally the following condition:

**COMPLETENESS** – Any objective property of a physical system *S* must be represented within the physical theory that is supposed to describe *S*.

It follows that there exist properties of physical systems that, according to the **REALITY** condition are objective – like [**spin**$_n$ = −1] for $S_2$ – but that QM does not represent as such: therefore QM is not complete.

As should be clear from the above presentation, the counterfactual reasoning here concerns the question: "Might the objective [**spin** $_n$ = −1] of $S_2$ have been somehow "created" by the spin measurement on the distant system $S_1$?". But *given* **LOCALITY**, the only acceptable possibility for justifying the existence of the property [**spin** $_n$ = −1] for the system $S_2$ also at a time $t'$ is its pre-measurement definiteness, but such definiteness is *entailed* by **LOCALITY** and nowhere the argument needs to *assume independently* that spin properties are definite. This is equivalent to stress that assuming **REALITY** is not equivalent to assuming at the outset the existence of definite, pre-measurement properties: **REALITY** is simply a sufficient criterion for a property of physical system to be objective, namely holding in a measurement-independent way[11].

---

[11] Maudlin goes as far as claiming that the EPR criterion – what I call here **Reality** – is an *analytic* criterion, namely a criterion that simply follows from the very meanings of the words that compose it (Maudlin 2014, 424010-6).

# 5  Conclusions

Alongside with a wide consensus on the relevance of the EPR-like arguments and the Bell theorem to deep foundational issues concerning fundamental physics, it is interest to note that a different, 'deflationary' attitude toward these issues is alive, shared in the past also by physicists of such high profile as Richard Feynman, Murray Gell-Mann, Jim Hartle. In a lecture held in 1983 Feynman claimed that the Bell theorem "is not a theorem that anybody thinks is of any particular importance. We who use quantum mechanics have been using it all the time. It is not an important theorem. It is simply a statement of something that we know is true – a mathematical proof of it." (quoted in Whitaker 2016, 493), while Murray Gell-Mann and Jim Hartle wrote:

> The EPR or EPRB situation is no more mysterious. There, a choice of measurements, say $\sigma_x$ or $\sigma_y$ for a given electron, is correlated with the behavior of $\sigma_x$ or $\sigma_y$ for another electron because the two together are in a singlet state even though widely separated. Again, the two measurement situations (for $\sigma_x$ and $\sigma_y$) decohere from each other. But here, in each, there is also a correlation between the information obtained about one spin and the information that can be obtained about the other. This behavior, although unfortunately called "non-local" by some authors, involves no non-locality in the ordinary sense of quantum field theory and no possibility of signaling outside the light cone. The problem with "local realism" that Einstein would have liked is not the locality but the *realism*. (Gell-Mann, Hartle 1989, 340).

A key strategy of particular expressions of this deflationary attitude lies in pointing out counterfactuality as the style of reasoning that carries with itself the allegedly 'realistic' burden of the Einstein-Bell arguments. In the above pages I have attempted to develop some arguments to counteract this attitude: if agreed upon, these arguments lead us to see that the real challenge is not the search for imaginary flaws in the Einstein-Bell work, but rather the exercise of scientific (and metaphysical!) imagination in order to figure out what a non-local natural world might look like.